\documentstyle[aps,prb,epsfig,twocolumn]{revtex}
\begin{document}

\draft

\twocolumn[
\hsize\textwidth\columnwidth\hsize\csname @twocolumnfalse\endcsname

\title{BCS theory for s+g-wave superconductivity in
borocarbides Y(Lu)Ni$_2$B$_2$C}
\author{Qingshan Yuan$^{1,2}$ and Peter Thalmeier$^1$}
\address{$^1$ Max-Planck-Institut f\"ur Chemische Physik fester Stoffe,
N\"othnitzer Str. 40, 01187 Dresden, Germany\\
$^2$Pohl Institute of Solid State Physics, Tongji University,
Shanghai 200092, P.R.China}

\maketitle

\begin{abstract}
The s+g mixed gap function $\Delta_{{\bf k}}=\Delta
{[(1-x)-x\sin ^4 \theta \cos 4\phi]}$ ($x$: weight of g-wave component)
has been studied within BCS theory. By suitable
consideration of the pairing interaction, we have confirmed that
the coexistence of s- and g-wave, as well as the state with equal s
and g amplitudes (i.e., $x=1/2$) may be stable.
This provides the semi-phenomenological theory for the
s+g-wave superconductivity with point nodes which has been
observed experimentally in borocarbides YNi$_2$B$_2$C and possibly in
LuNi$_2$B$_2$C.
\end{abstract}

]

\section{Introduction}

The rare earth nickel borocarbides RNi$_2$B$_2$C
(R=Y, Lu, Tm, Er, Ho, and Dy) have attracted
great interest in recent years due to superconductivity (SC) as well as
its possible coexistence with antiferromagnetic order.\cite{Canfield,Muller}
It has initially been thought that these materials can be understood
by an isotropic s-wave pairing via the conventional electron-phonon
coupling.\cite{Carter}
However, recent various experimental results particularly
on the two nonmagnetic borocarbides Y(Lu)Ni$_2$B$_2$C,
including specific heat,\cite{Nohara97,Nohara99,Izawa01,Park}
thermal conductivity,\cite{Izawa02,Boaknin} Raman scattering,\cite{Yang}
NMR relaxation rate,\cite{Zheng} photoemission spectroscopy,\cite{Yokoya}
scanning tunneling microscopy and spectroscopy,\cite{Samper}
have unambiguously shown that the gap function is highly anisotropic
with anisotropy ratio\cite{Izawa02} $\Delta_{min}/\Delta_{max}\leq$10$^{-2}$.
For example, the $\sqrt{H}$ dependence of the specific heat in the
vortex state indicates a superconducting state with
nodal excitations\cite{Nohara97,Nohara99,Izawa01}. The $T^3$ power law
behavior of the spin-lattice relaxation rate\cite{Zheng} also suggests
the existence of nodes. Very recently,
compelling evidence is presented by Izawa {\it et al.} from the
angular-dependent thermal conductivity in a magnetic field
that the gap function of YNi$_2$B$_2$C has point nodes which are
located along the [1,0,0] and [0,1,0] directions.\cite{Izawa02}
The same conclusion can be also drawn from the angular-dependent
specific heat data.\cite{Park}
Highly anisotropic s-wave gap (with possible nodes)
was also discovered in LuNi$_2$B$_2$C by thermal conductivity
measurements as a function of temperature and field
strength.\cite{Boaknin} Thus the previous isotropic s-wave theory has
to be critically reconsidered.

\begin{figure}[ht]
\begin{center}
\centerline{\epsfig{file=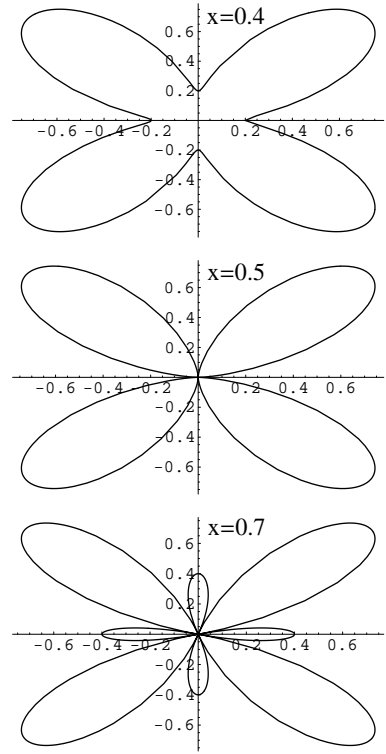,width=5cm,height=10cm,clip=}}
\medskip
\caption{The $xy$-plane ($\theta=\pi/2$) polar plots of the s+g wave
gap  $|\Delta_{\bf k}|$ for various tuning parameter $x$.}
\end{center}
\label{Fig:Plots}
\end{figure}

Recently Maki {\it et al.} have proposed that the so called s+g-wave
spin singlet gap function for Y(Lu)Ni$_2$B$_2$C superconductors, i.e.,
\cite{Maki,Thalmeier,Izawa02}
\begin{equation}
\Delta_{\bf k}={\Delta \over 2}(1 - \sin ^4\theta \cos 4\phi)\ ,
\label{gapfun0}
\end{equation}
is consistent with experimental results.
We introduced $\theta, \phi$ as polar and azimuthal angles of ${\bf k}$,
respectively. Here the second `g-wave' contribution is given by a
fourth degree fully symmetric (A$_{1g}$) basis function
$\psi^{(4)}(\theta,\phi)$ in tetragonal D$_{4h}$ symmetry which is, up to
a constant, equal to the real `tesseral harmonic'
$Z^c_{44}(\theta,\phi) =
(1/\sqrt{2})[Y_4^{4}(\theta,\phi)+Y_4^{-4}(\theta,\phi)]$
function. We have
\begin{eqnarray}
\psi^{(4)}(\theta,\phi)=k_x^4+k_y^4-6k_x^2 k_y^2=\sin^4\theta\cos4\phi \ .
\end{eqnarray}
In the gap function (\ref{gapfun0}),
the amplitudes of s- and g-components are assumed to be
equal. Thus 4 (and only 4) point nodes at
$\theta=\pi/2$ and $\phi=0, \pi/2, \pi, 3\pi/2$ are realized, see
the middle panel of Fig.~\ref{Fig:Plots}. This is
exactly what has been observed experimentally.\cite{Izawa02}

On the other hand,
there is no symmetry reason for the constraint of equal amplitudes of
s and g. More generally, the s+g gap function is:
\begin{equation}
\label{gapfun}
\Delta_{\bf k}=\Delta[(1-x)
- x \sin ^4\theta \cos 4\phi]=\Delta f(\theta,\phi)\nonumber
\label{gapfun}
\end{equation}
with a tuning parameter $x$ characterizing the weight of g-wave
component. Obviously Eq.~(\ref{gapfun0}) corresponds to the special
case $x=1/2$. If $x<1/2$, s-wave is dominant and the nodes will be
removed; while if $x>1/2$, g-wave has strong weight and will
contribute 8 line nodes. The three different cases have been shown
in $xy$-plane in Fig.~\ref{Fig:Plots}.

A natural question then arises how to understand the origin of the above
s+g hybrid pairing. So far, a microscopic theory for the pairing
potential is not available which might be complicated due to the
complex Fermi surface of borocarbides\cite{Dugdale99} and the
possibility of strongly anisotropic Coulomb interactions.
As a first step, however, it is necessary
to investigate phenomenologically how the s+g state can be realized by
constructing an appropriate pairing potential. This is the topic of the
present work. Within BCS weak-coupling theory we will show how a
stable coexistence of s- and g-wave and in particular how the fine
tuning $x=1/2$ (s+g) can be realized almost independent of temperature
below $T_c$.

A similar issue has been addressed by Lee and Choi\cite{Lee} in their
theory to explain Raman scattering data, but no complete analysis was
presented. First, they do not consider the $\theta$-dependence of the
g-wave part, i.e., implicitly assume a gap function with cylindrical symmetry.
Second, they use gap models with strong g-component. As shown
before in Fig. 1, these will exhibit 8 line nodes, which are
inconsistent with the experimental results\cite{Izawa02} which have
revealed 4 point nodes.

\section{Pairing interaction and BCS theory for the s+g  wave state}

In view of the orthogonality of s- and g-wave functions in Eq.~(\ref{gapfun}),
one may naturally express the pairing potential as sum of two
separable parts whose weight is given by two parameters $V_s$ and
$V_g$:\cite{Ghosh}
\begin{eqnarray}
V_{\bf kk'} & = &
-[V_s+V_g \psi^{(4)}(\theta,\phi) \psi^{(4)}(\theta',\phi')] \ .
\label{int0}
\end{eqnarray}
Here the unprimed and primed angles correspond to ${\bf k}$ and ${\bf
k'}$ respectively. Each term in Eq.~(\ref{int0}) is separable with
respect to wavevectors ${\bf k}$ and ${\bf k'}$.

Here we try to propose a pairing potential, similar to that
used in Ref.~\onlinecite{Lee} for the cylindrical gap,
by adding another mixing term $V_{sg}$:
\begin{eqnarray}
V_{\bf kk'} & = & -[V_s+V_g
\psi^{(4)}(\theta,\phi)\psi^{(4)}(\theta',\phi')\nonumber\\
&& \ \ \ +V_{sg}(\psi^{(4)}(\theta,\phi)+\psi^{(4)}(\theta',\phi'))]\ .
\label{int}
\end{eqnarray}
The reason for the choice of this type of interaction as well as
its derivation will be discussed later. We simply adopt it as the
model pairing potential for the moment.

The standard BCS gap equation is then written as
\begin{eqnarray}
\Delta_{\bf k} & = & -\sum_{\bf k'} V_{\bf kk'} {\Delta_{\bf k'}
\over 2E_{\bf k'}} \tanh (\beta E_{\bf k'} /2) \ ,\label{gapeqn}
\end{eqnarray}
where $\beta=1/(k_B T)$ and $E_{\bf k}=\sqrt{(\varepsilon_{\bf
k}-\mu)^2+\Delta_{\bf k}^2}$ is
the quasiparticle spectrum. $\varepsilon_{\bf k}$ is the free electron
dispersion, and $\mu$ is the chemical potential. It is easy to check
that the gap function (\ref{gapfun}) is a self-consistent solution
of Eq.~(\ref{gapeqn}) with inclusion of the pairing interaction (\ref{int}),
if the gap amplitude $\Delta$ and tuning parameter $x$ satisfy the
following self-consistent equations:
\begin{eqnarray}
1-x & = & V_s \sum_{\bf k} {f(\theta_{\bf k},\phi_{\bf k}) \over
2E_{\bf k}} \tanh (\beta E_{\bf k} /2)\nonumber\\
& &  +V_{sg} \sum_{\bf k}\psi^{(4)}(\theta_{\bf k},\phi_{\bf k})
{f(\theta_{\bf k},\phi_{\bf k}) \over 2E_{\bf k}}
\tanh (\beta E_{\bf k} /2) \ ,\\
-x & = & V_g \sum_{\bf k}\psi^{(4)}(\theta_{\bf k},\phi_{\bf k})
{f(\theta_{\bf k},\phi_{\bf k}) \over 2E_{\bf k}}
\tanh (\beta E_{\bf k} /2)\nonumber\\
& & +V_{sg} \sum_{\bf k} {f(\theta_{\bf k},\phi_{\bf k}) \over
2E_{\bf k}} \tanh (\beta E_{\bf k}/2) \ .
\end{eqnarray}
where the angles have been indexed by their corresponding wavevector
for clarity. Replacing the summation by integration according to
$$
\sum_{\bf k}\simeq {N(0) \over 4\pi}
\int_{-\hbar \omega_D}^{\hbar \omega_D} d\varepsilon \int d\Omega \ ,
$$
where $\hbar \omega_D$ is an energy cut-off to enforce the
constraint $|\varepsilon_{\bf k}-\mu|\le \hbar \omega_D$ ($\omega_D$: Debye
frequency for phonon-mediated SC), and $N(0)$ is the density of
states at zero energy for the spectrum $\varepsilon_{\bf k}-\mu$,
we may obtain the following equations:
\begin{eqnarray}
1-x & = & {1\over 4\pi} (\tilde{V}_s I_1 + \tilde{V}_{sg} I_2)\ ,
\label{eq:Vs}\\
-x & = & {1\over 4\pi} (\tilde{V}_g I_2 + \tilde{V}_{sg} I_1)\ .
\label{eq:Vg}
\end{eqnarray}
Above, $\tilde{V}_s=N(0)V_s,\ \tilde{V}_g=N(0)V_g,\
\tilde{V}_{sg}= N(0)V_{sg}$ are redefined dimensionless interaction constants,
and the integrals $I_{1,2}$ are written as follows
\begin{eqnarray}
I_1 & = & \int_{0}^{1} d\varepsilon \int d\Omega
{f \tanh (\beta \sqrt{\varepsilon^2+\Delta^2f^2}/2)\over
\sqrt{\varepsilon^2+\Delta^2f^2}} \ ,\\
I_2 & = & \int_{0}^{1} d\varepsilon \int d\Omega \psi^{(4)}
{f \tanh (\beta \sqrt{\varepsilon^2+\Delta^2f^2}/2)\over
\sqrt{\varepsilon^2+\Delta^2f^2}} \ ,
\end{eqnarray}
where we use the abbreviated symbols $f$ and $\psi^{(4)}$, and $\hbar \omega_D$
has been taken as the energy unit.

\section{Numerical results}

\begin{figure}[ht]
\centerline{\epsfig{file=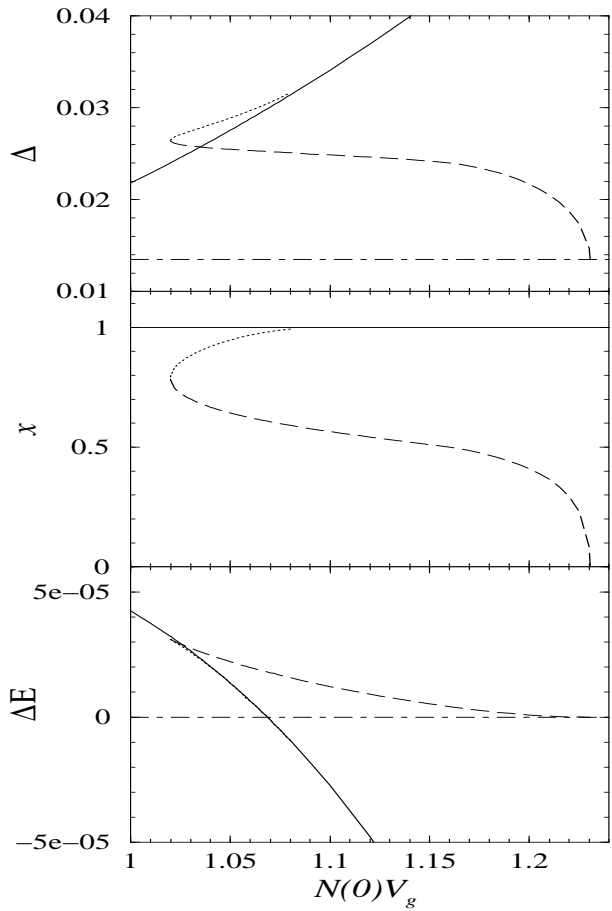,width=8cm,height=12cm,clip=}}
\smallskip
\caption{The values $\Delta$, $x$, and relative energies $\Delta$E
as functions of $\tilde{V}_g$ for $\tilde{V}_s=0.2$ and $T=0$. The
solid and dot-dashed lines are for pure g- and s-wave solutions,
respectively. The dotted and dashed lines are for two possible s+g
mixed solutions. The energy of the pure s-wave solution is taken
as the reference point in the lowest panel. Energy unit for
$\Delta$ is $\hbar\omega_D$ and for $\Delta$E is
$N(0)(\hbar\omega_D)^2$.} \label{Fig:sgT0}
\end{figure}

We first consider $V_{sg}=0$, i.e., we assume the pairing interaction
(\ref{int0}). It was found that one or two s+g solutions
(i.e., $\Delta>0,\ 0<x<1$) may appear when $\tilde{V}_g$ is a few times
greater than $\tilde{V}_s$. On the other hand, it is easy to check
that the pure s-wave ($x=0$) and pure g-wave ($x=1$) are always
trivial solutions.
Thus one needs to compare their free energies to find
the stable solution. In unit of $N(0)(\hbar \omega_D)^2$ the free
energy is given by
\begin{eqnarray*}
F & = & {-1\over 2\pi}\int_{0}^{1} d\varepsilon \int d\Omega
[\sqrt{\varepsilon^2+\Delta^2f^2}+{2\over \beta}
\ln (1+e^{-\beta\sqrt{\varepsilon^2+\Delta^2f^2}})] \nonumber\\
& & +\Delta^2 (1-x)^2/\tilde{V}_s + \Delta^2 x^2/\tilde{V}_g\ .
\end{eqnarray*}
Detailed calculation shows that the s+g mixed state is unstable
in most of the parameter space.
As an example, we have shown in Fig.~\ref{Fig:sgT0}
all the solutions and their relative energies as functions of
$\tilde{V}_g$ for fixed $\tilde{V}_s=0.2$ and $T=0$.
Two s+g solutions may be present, as shown by the dotted and dashed lines.
But compared to the pure s- and/or g-wave solutions, they are found to be
energetically unfavorable, see the lowest panel in
Fig.~\ref{Fig:sgT0}.\cite{Supp}
Thus the s+g mixture seems very unlikely under the pairing
interaction which is the sum of two separable parts (\ref{int0}).

Once $V_{sg}>0$, the above situation changes substantially. In this
case the pure s- and g-wave are no longer solutions of Eqs.~(\ref{eq:Vs})
and (\ref{eq:Vg}). Furthermore only one stable s+g solution is present.
We explain the details in the following.

\begin{figure}[ht]
\centerline{\epsfig{file=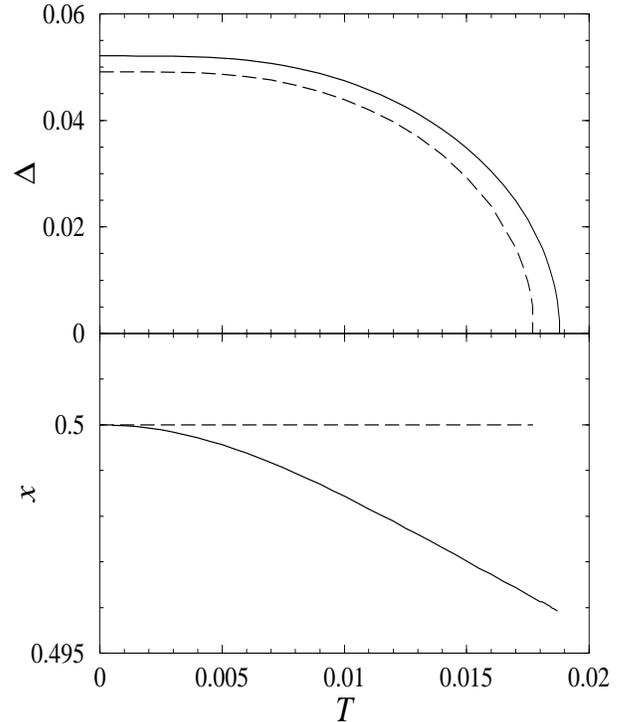,width=8cm,height=10cm,clip=}}
\smallskip
\caption{The order parameter $\Delta$ and tuning parameter $x$ as
functions of temperature $T$. $\Delta$ and $T$ are in units of
$\hbar\omega_D$. The solid
lines are for $\tilde{V}_s=0.2,\ \tilde{V}_g=0.1,\
\tilde{V}_{sg}=-0.22$, and the dashed lines are for
$\tilde{V}_s=\tilde{V}_g=-\tilde{V}_{sg}=0.2$ corresponding to the
special case of $T$-independent $x=1/2$ s+g wave.}
\label{Fig:OP}
\end{figure}

First we consider a special case: $V_s=V_g=-V_{sg}$. Then,
by adding Eqs.~(\ref{eq:Vs}) and (\ref{eq:Vg}) one can immediately
obtain the solution with $x=1/2$ independent of temperature. This
means that irrelevant of
temperature s- and g-wave coexists always with equal amplitudes.
This result is obvious because in this case the pairing interaction
(\ref{int}) can be simply factorized  again into the form
$f(\theta,\phi)*f(\theta',\phi')$ with fixed $x=1/2$.
Then only the gap amplitude $\Delta(T)$ is left. It decreases
gradually with $T$ and vanishes at the transition temperature $T_c$,
as shown by the dashed line in the upper panel of Fig.~\ref{Fig:OP}
where $\tilde{V}_s=0.2$ is used.

Generally, the three interaction parameters have different absolute
values. Various situations can be described by tuning these model parameters.
We use the following strategy to obtain reasonable values:
First, we fix the value of $\tilde{V}_s$, e.g. $0.2$ throughout the
work which sets the overall scale for $T_c$. Then we assume a value
for $\tilde{V}_g$ smaller than  $\tilde{V}_s$ and tune
$\tilde{V}_{sg}$ to realize the s+g state with the constraint: $x=1/2$
at $T=0$. Experimentally, the detection of nodes by the field
angular-dependent thermal conductivity is applicable only at very low
temperatures, i.e., $T\ll T_c$. Thus the measurement actually provides
this constraint on the gap function.
In this way, the value $\tilde{V}_{sg}$
may be determined for each given $\tilde{V}_g$. 
We have obtained a nearly linear relation
between $\tilde{V}_{sg}$ and $\tilde{V}_g$: 
$\tilde{V}_{sg}\simeq -0.24+0.2\tilde{V}_g$.
With interaction parameters fixed we can now study the intriguing
issue how the $x=1/2$ fine tuning s+g state at $T=0$
evolves with temperature. One would expect that
not only the gap amplitude $\Delta$, but also
the tuning parameter $x$ will change with temperature. If it would
decrease, the node points would cease to exist and a gap would open
with increasing temperature. In principle this is indeed observed.
As an example, we take $\tilde{V}_g=0.1<\tilde{V}_s$.
Then $\tilde{V}_{sg}\simeq -0.22$ is obtained to realize the $x=1/2$ s+g
solution at $T=0$. Under these interaction parameters,
$\Delta(T)$ and $x(T)$ are calculated self-consistently from
Eqs.~(\ref{eq:Vs}) and (\ref{eq:Vg}). The results are shown by the solid
lines in Fig.~\ref{Fig:OP}. It is interesting to see that $x$ varies with $T$
monotonically and very slowly. In the current example with $V_g<V_s$,
$x$ becomes less than $1/2$ at finite $T$. However, the deviation from $1/2$
is less than $1\%$ even at $T=T_c$. This means that
very strong anisotropic gap with
$\Delta_{min}/\Delta_{max}\leq$10$^{-2}$ is present in the whole
superconducting region describing essentially still a gap with point nodes.
Thus we conclude that s+g pairing with nodal excitations is a robust
solution for all temperatures and should not be considered as accidental.
We also mention that $x$ becomes larger than $1/2$ at finite $T$
for $V_g>V_s$ (not shown), but again only a small deviation is obtained.

\section{Discussion and conclusion}

In the above section, we have obtained the stable s+g-wave
based on the phenomenological pairing interaction (\ref{int}). Also
we have realized the $x=1/2$ hybrid state which has point nodes
by appropriate choice of the interaction parameters, and most
importantly have proven its robustness with temperature variation.
A microscopic justification for our phenomenological scenario is yet
to be investigated. On the other hand, we should return back to
understand the form of the pairing potential Eq.~(\ref{int}) in more
detail. This is discussed in the following.

For a system with full translation symmetry, the pairing potential
$V_{\bf kk'}$ can only be a function of ${\bf k}-{\bf k'}$. 
Further it may be considered to be a function of the angle between the
two wavevectors in view of $|{\bf k}|,|{\bf k'}|\simeq k_F$ 
(the Fermi wave vector). Then $V_{\bf kk'}=V(\hat{\bf k}\cdot\hat{\bf k'})$, 
with $\hat{\bf k}\cdot\hat{\bf k'}$ denoting the cosine of the angle between 
${\bf k}$ and ${\bf k'}$, can be expanded in terms of Legendre polynomials. 
And by use of the spherical harmonic addition theorem, 
it can be finally written into
\begin{equation}
V(\hat{\bf k}\cdot\hat{\bf k'}) = \sum_{l=0}^{\infty}V_l \sum_{m=-l}^{l} Y_{lm}
(\theta,\phi) Y_{lm}^*(\theta',\phi')\, \label{Vkk'}
\end{equation}
through spherical harmonics $Y_{lm} (\theta,\phi)$. Obviously the
$V_s$ term corresponds to $l=0$ and the $V_g$ term results from the
sum of $(l,m)=(4,4)$ and $(4,-4)$. However the cross term $\sim V_{sg}$
in Eq.~(\ref{int}) cannot be obtained from Eq.~(\ref{Vkk'}). This leads us
to go beyond
the above assumption, i.e., to consider $V_{\bf kk'}$ as a general
function which depends on ${\bf k}$ and ${\bf k'}$ individually. This
will be the case when the effect of having only discrete
lattice translation and rotation symmetry is included. Then the pair
potential has to be expanded in terms of the basis functions
$\psi^{i(l)}_\Gamma(\theta,\phi)$ of the crystal symmetry group
(D$_{4h}$) belonging to a specific irreducible representation $\Gamma$ of
degree $l$ and degeneracy index $i$. The generalized expansion then reads,
suppressing the multiplicity index of $\Gamma$:
\begin{equation}
V_{\bf kk'} = \sum_{\Gamma ll'}V^{(ll')}_\Gamma(k,k')\sum_i
\psi^{i(l)}_\Gamma(\theta,\phi)\psi^{i(l')}_\Gamma(\theta',\phi')^*.
\end{equation}
For basis functions of different degree $l,\ l'$ but belonging
to the same representation $\Gamma$ the contribution will
generally be nonzero. Then for $\Gamma$= A$_{1g}$ and $l=0,\ l'=4$
or vice versa one can naturally obtain the nondiagonal
contributions $V^{(04)}_{{\rm A}_{1g}}=V_{sg}$ in D$_{4h}$ symmetry.

In conclusion, the s+g hybrid gap function has been studied within BCS theory.
By an appropriate ansatz for the pairing interaction and selection of
the interaction parameters we have shown that the stable coexistence
of s- and g-wave, as well as the nodal state with equal amplitudes of
them is possible. In particular, we have confirmed that the hybrid state with
highly anisotropic gap may be robust in the whole superconducting region.
This provides the semi-phenomenological theory to understand the
s+g $(x=1/2)$ gap function proposed by Maki {\it et al.}\cite{Maki}
Thus the explanation of the experimentally observed point nodes
in borocarbides Y(Lu)Ni$_2$B$_2$C can be achieved
in a self-consistent way. Finally we mention that similar
s+g mixed gap functions are also proposed very recently for skutterudite
PrOs$_4$Sb$_{12}$,\cite{Maki03} and their justification
should be possible based on a similar analysis as presented here.

\section*{acknowledgment}
We thank J. Goryo, K. Maki, and Q. Gu for useful discussions. Q. Yuan also
acknowledges the support by the National Natural
Science Foundation of China (Grant No. 19904007).


\begin{references}
\bibitem{Supp} Only in a narrow region about $1.07<\tilde{V}_g<1.08$,
the dotted line, which is indiscernible from the solid one, has the lowest
energy, indicating a stable s+g solution with $x\approx 1$.
\end{references}
\end{document}